\documentclass[a4paper,fleqn,usenatbib]{mnras}

\usepackage{graphicx}	
\usepackage{amsmath}	
\usepackage{amssymb}	

\usepackage[T1]{fontenc}
\usepackage{ae,aecompl}

\usepackage{txfonts}

\hypersetup{draft}

\newcommand{\bc}{\begin{centre}}
\newcommand{\ec}{\end{centre}}

\newcommand{\Mpc}{~h^{-1}~{\rm Mpc}}

\newcommand{\Hunit}{~h~{\rm km}~s^{-1}~{\rm Mpc}^{-1}}

\title[The baryonic cosmic web]{The large-scale environment from cosmological simulations I: The baryonic cosmic web}

\author[Weiguang Cui et al.]
{\parbox{\textwidth}{Weiguang Cui,$^{1}$\thanks{E-mail: \texttt{weiguang.cui@uam.es}}
 Alexander Knebe,$^{1,2}$ Gustavo Yepes,$^{1,2}$ Xiaohu Yang,$^{3,4}$ Stefano Borgani,$^{5,6,7}$
 Xi Kang,$^8$ Chris Power,$^{9,10}$ Lister Staveley-Smith$^{9,10}$.}\vspace{0.4cm}
 \\
 \parbox{\textwidth}{
  $^1$ Departamento de F\'isica Te\'{o}rica, M\'{o}dulo 15, Facultad de Ciencias,
  Universidad Aut\'{o}noma de Madrid, 28049 Madrid, Spain\\
  $^2$ Astro-UAM, UAM, Unidad Asociada CSIC\\
  $^3$ Department of Astronomy, Shanghai Jiao Tong University,
  Shanghai 200240, China\\
  $^4$ IFSA Collaborative Innovation centre, Shanghai Jiao Tong University,
  Shanghai 200240, China\\
  $^5$ Astronomy Unit, Department of Physics, University of Trieste,
  via Tiepolo 11, I-34131 Trieste, Italy\\
  $^6$ INAF -- Astronomical Observatory of Trieste, via Tiepolo 11,
  I-34131 Trieste, Italy\\
  $^7$ INFN -- Sezione di Trieste, I-34100 Trieste, Italy\\
  $^8$ Purple Mountain Observatory, the Partner Group of MPI f\"ur Astronomie,
  2 West Beijing Road, Nanjing 210008, China\\
  $^9$ International Centre for Radio Astronomy Research (ICRAR),
  University of Western Australia, 35 Stirling Highway,
  Crawley, Western Australia 6009, Australia\\
  $^{10}$ ARC Centre of Excellence for All-Sky Astrophysics (CAASTRO)\\
}}

\date{Accepted XXX. Received YYY; in original form ZZZ}

\pubyear{2017}

\begin{document}
\label{firstpage}
\pagerange{\pageref{firstpage}--\pageref{lastpage}}
\maketitle

\begin{abstract}
Using a series of cosmological simulations that includes one dark-matter-only (DM-only) run, one gas cooling-star formation-supernovae feedback (CSF) run and one that additionally includes feedback from active galactic nuclei (AGNs), we classify the large-scale structures with both a velocity-shear-tensor code ({\sc Vweb}) and a tidal-tensor code ({\sc Pweb}). We find that the baryonic processes have almost no impact on large-scale structures -- at least not when classified using aforementioned techniques. More importantly, our results confirm that the gas component alone can be used to infer the filamentary structure of the Universe practically un-biased, which could be applied to cosmology constrains. In addition, the gas filaments are classified with its velocity ({\sc Vweb}) and density ({\sc Pweb}) fields, which can theoretically connect to the radio observations, such as $H_I$ surveys. This will help us to bias-freely link the radio observations with DM distributions at large scale.
\end{abstract}



\section{Introduction}
\label{i}

On large scales, matter in the Universe can be roughly distributed into knots, filaments, sheets and voids, which form the rather prominent cosmic web seen both in numerical simulations of cosmic structure formation and observations of the distribution of galaxies. These four different cosmological structures are a natural outcome of gravitational collapse. A detailed understanding of the large-scale environment (LSE) helps us to model both how the DM or galaxies are distributed and evolve from early times to the present day. It has been shown that there is an interplay between the LSEs and galaxies residing in them in terms of, for instance, star formation \citep[e.g.][]{Peng2012, Darvish2017, Kuutma2017}, fractions of red galaxies \citep[e.g.][]{Wang2016a}, orientation \citep[e.g.][]{Zhang2013a} and spin direction \citep[e.g.][]{Zhang2015, Pahwa2016}. The DM haloes hosting these galaxies are also affected by the large-scale structure of the Universe with respect to, for instance, mass, shapes, and formation times \citep[e.g.][]{Hahn2006, Lee2008, Metuki2014}, spin and orientation directions \citep[e.g.][]{Aragon-Calvo2007, Zhang2009}, peculiar velocity profiles \citep[e.g.][]{Lee2016}, and halo bias \citep{Yang2017}. And the LSE might also hold the clue to understanding peculiarities observed in the substructure content of haloes such as, for instance, the observed planar distribution of satellite galaxies \citep[e.g.][]{Libeskind2015}, satellite alignments \citep[e.g.][]{Chen2015, Tempel2015}, and spins and mergers \citep[e.g.][]{Tempel2013, Dubois2014, Welker2014, Kang2015}. Precise modelling and understanding of galaxy formation therefore requires an exquisite comprehension of the cosmic web, too.

Many methods have been developed to classify/identify these cosmological structures. From a theoretical point of view, a classification scheme normally splits (either configuration or velocity) space into cells. Then, the eigenvalues of the Hessian matrix $\mathbfit{T}_{\alpha\beta}$ for the indicator field are used to separate out these structures spatially. This tracer normally uses a smoothed (logarithmic) density field \citep[{\sc Tweb}][]{Aragon-Calvo2010b}; \citep[{\sc NEXUS}][]{Cautun2012a}) or the tidal field of the gravitational potential ({\sc Pweb}, \cite{Hahn2006}, and its extensions, \cite{Forero-Romero2009}), density field \citep{Zhang2009}, velocity shear tensor ({\sc Vweb}, \cite{Hoffman2012}, also \cite{Cautun2012a}, and its particle-based formulation, \cite{Fisher2016}) or velocity divergence \citep{Cautun2012a}, and shear of the Lagrangian displacement field \citep[{\sc DIVA},][]{Lavaux2010}. The identification scheme includes {\sc SpineWeb} \citep[invokes local adjacency properties of the boundaries between the watershed basins to trace the critical points in the density field][]{Aragon-Calvo2010}, {\sc DisPerSE} \citep[based on the discrete Morse theory][]{Sousbie2011a, Sousbie2011b}, and {\sc ORIGAMI} \citep[based on the number of orthogonal axes along which stream-crossing occurs][]{Falck2012}. There are more methods that only focus on one particular structures, for example, using the multi-stream field based on velocity flows \citep{Shandarin2011}, on tessellations of density field \citep{Shandarin2011}, flip-flop field in Lagrangian space \citep{Shandarin2016}, the path density method \citep{Genovese2009} or using the marked point process \citep[based on a stochastic method with Markov-chain Monte Carlo algorithm, Bisous model][]{Tempel2014b, Tempel2016} to identify cosmic webs, using density depressions \citep[{\sc ZOBOV},][]{Neyrinck2008} or the Watershed method \citep[{\sc WVF},][]{Platen2007} to identify voids. We refer to \cite{Cautun2012a, Leclercq2016, Libeskind2017} for comparisons between these structure identification methods.

In the literature, dark-matter-only simulations are normally used with aforementioned methods to study the properties of the large-scale structure -- which is dominated by the effects of gravity and hence DM. It is well known that baryons impact upon structure formation on small, non-linear scales \citep[for example][and the references therein]{Cui2016a}, but it might have little impact on large-scales \citep[see a review in][]{CuiBC2017}. It therefore remains unclear whether baryonic (and especially gaseous) filaments and sheets will form (and/or follow) the same structures as DM. Here we aim at directly addressing cosmic web classifications and its relation to baryonic physics; we are simply asking the questions {\em do baryons affect these cosmological structures?} and {\em do baryons follow the (dark) matter cosmic web?}

On the observational side, different techniques are applied to classify and quantify LSE as many of aforementioned codes are primarily designed for cosmological DM simulations and hence cannot be directly applied to observational surveys. For instance, gas (and chiefly $H_I$) will be a very important tracer of the cosmic web, especially with the next generation radio telescope such as the already operational Australian Square Kilometre Array Pathfinder (ASKAP) and the coming Square Kilometre Array (SKA). But will the observed gas distribution be an un-biased tracer of the cosmic web? Galaxies, for instance, only provide a biased view of the large-scale structure distribution \citep[for example][]{Mo1997, Benson2000, Yang2012, Wang2013}. They are nevertheless used to reconstruct filaments by estimating a smoothed density field Hessian matrix based on their spatial distribution \citep[e.g.][]{Darvish2017}, by connecting individual segments that are found in galaxy distribution \citep[the Candy model][]{Stoica2005, Zhang2009}, or by applying the Subspace Constrained Mean Shift algorithm \citep[e.g.][]{Chen2015}.

In the following sections, we briefly describe the full-physics hydrodynamical simulations used here and the different sub-grid models (\S~\ref{simulation}), and present the two cosmological structure classification methods (\S~\ref{method}). In Section~\ref{results} we present our results. Finally, we summarize our conclusions in \S~\ref{concl}, and comment on the applications for interpretation of observations.

\begin{figure*}
 \includegraphics[width=0.9\textwidth]{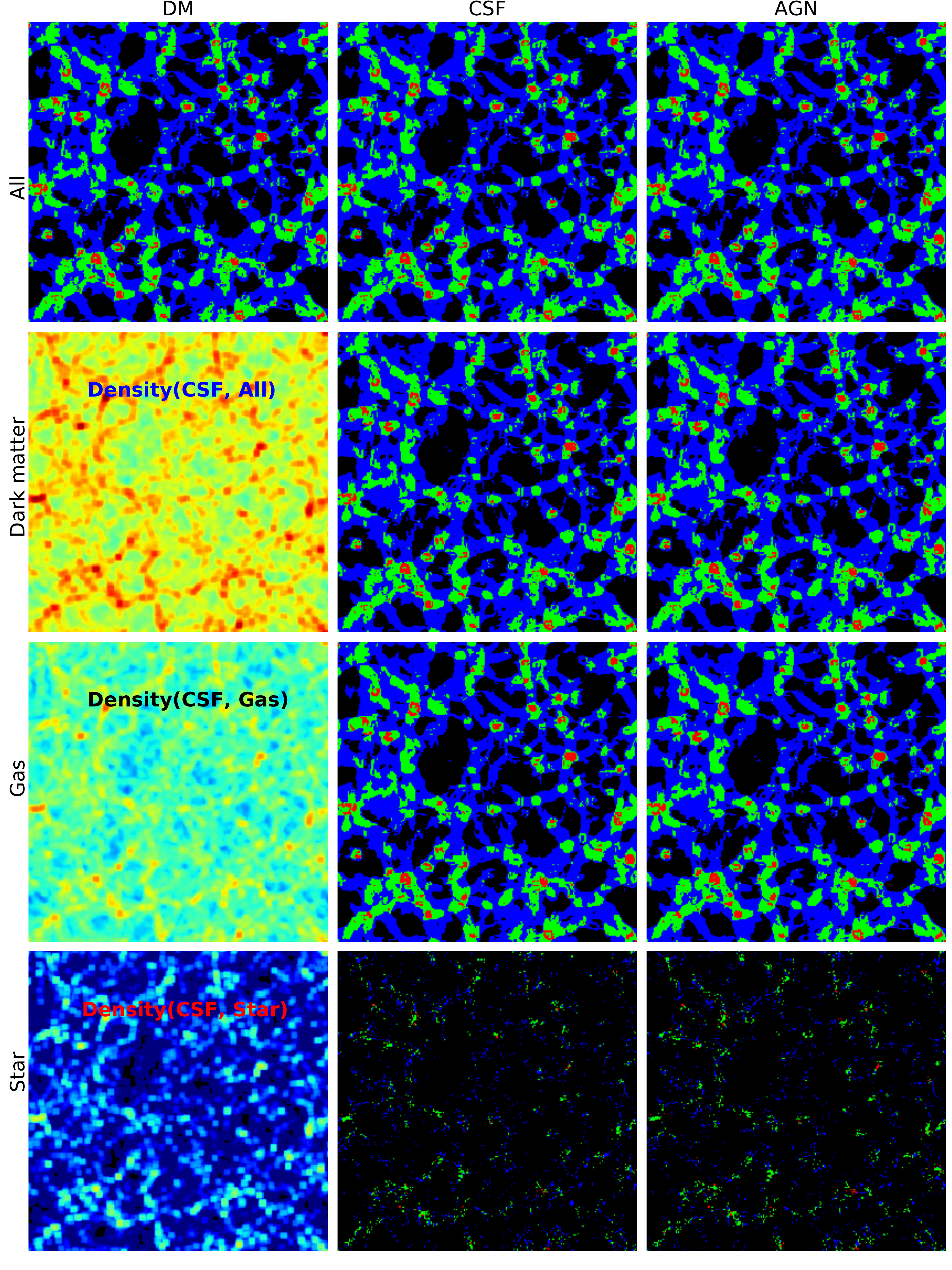}
 \caption{
  The projected structures for a slice ($\sim 1.6 \Mpc$, grid size) of the simulations from the {\sc Vweb} method: knots, filaments, sheets and void regions, which are shown as red, green, blue and black colours, respectively. The three columns from left- to right-hand side show the results from the DM-only, CSF and AGN runs. The classification results with all matter, DM, gas and stellar components are shown from the top to bottom rows, respectively. The left-hand lower three panels show the projected all matter, gas and stars densities from the CSF run, respectively.}\label{fig:show_v}
\end{figure*}
\begin{figure*}
 \includegraphics[width=0.9\textwidth]{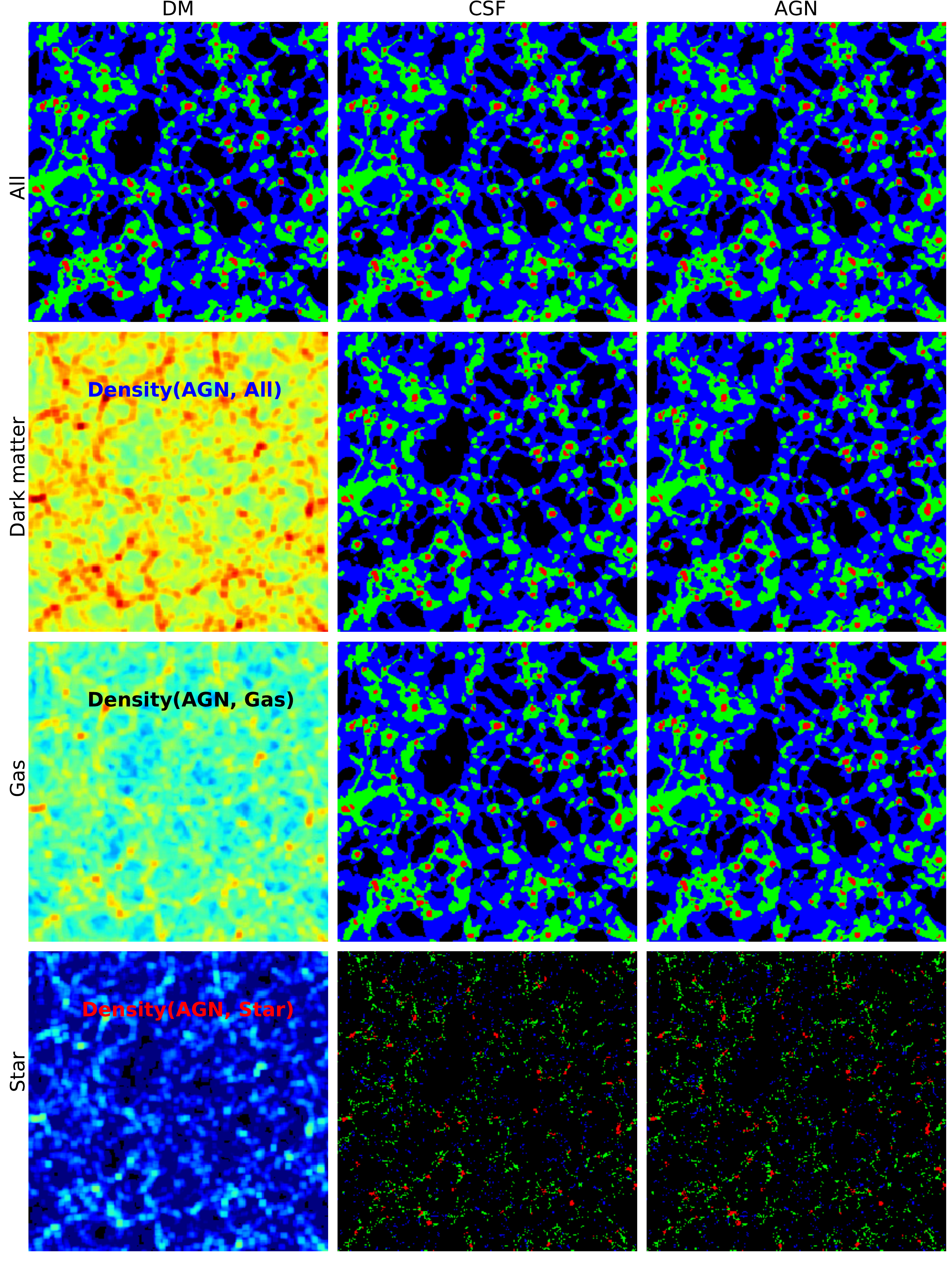}
 \caption{
  Same as Fig.~\ref{fig:show_v}, but for the {\sc Pweb} method. But please note that three lower panels in the first column now show the logarithmic densities in the AGN run for the various components.}\label{fig:show_p}
\end{figure*}

\section{The simulations}
\label{simulation}

A series of three cosmological simulations with boxsize $410 \Mpc$ are used in this paper to study the LSE. These simulations use a flat $\Lambda$ cold dark matter cosmology, with cosmological parameters of $\Omega_{\rm m} = 0.24$ for the matter density parameter, $\Omega_{\rm b} = 0.0413$ for the baryon contribution, $\sigma_8=0.8$ for the power spectrum normalization, $n_{\rm s} = 0.96$ for the primordial spectral index, and $h =0.73$ for the Hubble parameter in units of $100 \Hunit$. They used the same realization of the initial matter power spectrum, and were run with the TreePM-SPH code {\sc GADGET-3}, an improved version of the public {\sc GADGET-2} code \citep{Gadget2}. We refer to the dark-matter-only simulation as the DM-only run. The two hydrodynamical simulations both include radiative cooling, star formation and kinetic feedback from supernovae. In one case, we ignore feedback from AGN (which is referred to as the CSF run), while in the other we include it (which is referred to as the AGN run). The DM-only run has two families of DM particles: the one with larger particle mass shares the same ID as the DM particles in the CSF and AGN runs; while the one with smaller particle mass has equal mass to the gas particles in the CSF and AGN runs at the initial condition of $z$ = 49. With this particular setup, we can investigate the effect of baryons in detail. Interesting readers are referred to \cite{Cui2012a, Cui2014a} for details of these simulations and to \cite{Cui2016b, Cui2017a} for the statistical samples of galaxy clusters.

\section{The environment classification Methods --- {\sc Vweb} and {\sc Pweb}}
\label{method}

In this paper, we use both the {\sc Vweb} and the {\sc Pweb}\footnote{Please note that the {\sc Pweb} method at here is also referred as {\sc Tweb} in the previous literatures. It is not the same as the {\sc P-web} method in \cite{Tempel2014a}.} technique to classify the LSE of the three simulations. Both codes are particularly helpful for the investigations in this paper. Because the two methods use both velocity ({\sc Vweb}) and gravitational potential ({\sc Pweb}) information to quantify the LSE, these two quantities are directly connected with the radio emissions from gas filaments, which could be observed by the upcoming SKA telescopes \citep{Popping2015, Vazza2015}.

For the {\sc Vweb} method, the velocity shear tensor is used to classify the large-scale space into different structures. Following \cite{Hoffman2012}, it is defined as
\begin{equation}
 \Sigma_{\alpha\beta} = - \frac{1}{2H_0} \left( \frac{\partial v_\alpha}{\partial r_\beta} + \frac{\partial v_\beta}{\partial r_\alpha} \right),
\end{equation}
where, $H_0$ is the Hubble constant. The eigenvalues of $\Sigma_{\alpha\beta}$ are denoted as $\lambda^V_i$ ($i$ = 1, 2 and 3).

For the {\sc Pweb} method, the Hessian matrix of the gravitational potential field is used to classify these large-scale structures. Following \cite{Hahn2006}, it is defined as
\begin{equation}
 P_{\alpha\beta} = \frac{\partial^2 \Phi}{\partial r_\alpha \partial r_\beta},
\end{equation}
where $\Phi$ is the gravitational potential and the eigenvalues of $P_{\alpha\beta}$ are denoted as $\lambda^P_i$ ($i$ = 1, 2 and 3).

The computation of the eigenvalues for both matrices was performed on regular $256^3$ grids, corresponding to a cell size of $\sim 1.6 \Mpc$. We use a triangular-shaped cloud (TSC) prescription for the assignment of the particles and then compute the gravitational potential and the eigenvalues of the velocity shear tensor for every grid cell. In addition, we smoothed the quantities inside each cell to a scale of $\sim 8 \Mpc$ with a simple top-hat kernel. It has been shown in \cite{Tempel2014a} that the {\sc Vweb} filaments are independent of the smoothing scale. Therefore, we expect that the smoothing scale leaves little effect on our results.

Each individual cell is then classified as either `void', `sheet', `filament', or `knot' according to the eigenvalues $\lambda_1 > \lambda_2 > \lambda_3$ as follows:
\begin{itemize}
 \item[1.] void, if $\lambda_1 < \lambda_{th}$,
 \item[2.] sheet, if $\lambda_1 \geq \lambda_{th} > \lambda_2$,
 \item[3.] filament, if $\lambda_2 \geq \lambda_{th} > \lambda_3$, and
 \item[4.] knot, if $\lambda_3 \geq \lambda_{th}$,
\end{itemize}
where $\lambda_{th}$ is a free threshold parameter \citep{Hoffman2012, Libeskind2012, Libeskind2013}. Following the discussion of \cite{Hoffman2012, Carlesi2014}, we set $\lambda^V_{th} = 0.1$ for the {\sc Vweb} code. To mimic the LSE from {\sc Vweb} result, we find that $\lambda^P_{th} = 0.01$ is a suitable value for {\sc Pweb}. This value is very close to 0, which had been adopted by \cite{Hahn2006}. In addition, as the gas and DM share the same velocity field, the same threshold ($\lambda^V_{th} = 0.1$) is used for the gas component in the {\sc Vweb} code. However, the potential field obtained from only the gas component is lower than from DM (see density field in Figs.~\ref{fig:show_v} and \ref{fig:show_p} for details). Therefore, a even smaller threshold ($\lambda^P_{th} = 0.01 * \Omega_b/\Omega_m$) is applied to the {\sc Pweb} code when calculating it for the gas component. Resolution effects in the web classification are discussed in the Appendix \ref{A:denspeak} and have little impact on the results presented in this paper.

\section{Results}
\label{results}

\subsection{{\sc Vweb} versus {\sc Pweb} classification}
In Fig.~\ref{fig:show_v}, we illustrate how these {\sc Vweb} classified cosmological structures are distributed in these simulations. They are shown by different colours: red regions for knots, green regions for filaments, blue regions for sheets and black regions for voids. These plots are projections of a slice in $z$ of thickness one cell into the 2D $xy$-plane for the whole simulation boxsize ($410 \Mpc$). From left- to right-hand side, the columns show the results from DM, CSF and AGN runs. Besides the classification from using all types of particles (top row), we also show the cosmology structures classified by individual particle types --- DM (second row), gas (third row), and stars (fourth row). As these plots are either redundant or not available for the DM-only run, we show in the three panels under the DM-only run: the projected logarithmic densities of all matter (second row), gas (third row) and stars (last row) as found in the CSF run.

The first impression in Fig.~\ref{fig:show_v} is that the {\sc Vweb} code does provide in general a faithful classification of the cosmological structures. The void regions are surrounded by sheets; the filaments connect sheets and interconnect at knots. We find that the DM certainly dominates the classification, but we also confirm that the gas component alone reproduces nearly the same classification. The stars, however, seem not proper to be used to define a cosmic web for {\sc Vweb}-- at least not the same as found in the gas and/or dark-matter component. This could be cased by the coarse velocity field sampled by only star particles in these simulations. The LSEs classified by stars with {\sc Pweb} code seem to have a better agreement with the LSEs from the other tracers. This could means that the stars/galaxies can be a better tracer of the density/potential field in real space than of the velocity field in phase space. However, in both cases, further smoothing of the mesh cells or additional calibration of the parameters are required to use the star particles as a tracer for the codes. We further note that the detailed feedback schemes leave little effect on the web classification.

Using the same ordering and logic for the panels as in Fig.~\ref{fig:show_v}, we show in  Fig.~\ref{fig:show_p} the results from  the {\sc Pweb} code. With the chosen thresholds for $\lambda^P$, the LSE is very similar to the results from the {\sc Vweb} code. The conclusions drawn from Fig.~\ref{fig:show_p} are basically identical to the ones obtained for the {\sc Vweb} analysis. Note that the {\sc Vweb} reveals finer details of the cosmic web as opposed to the {\sc Pweb}, even though this result depends on the particular choice of the thresholds $\lambda^V_{th}$ and $\lambda^P_{th}$.

\subsection{Baryon effects}
\label{4.0}
\begin{figure*}
 \includegraphics[width=\textwidth]{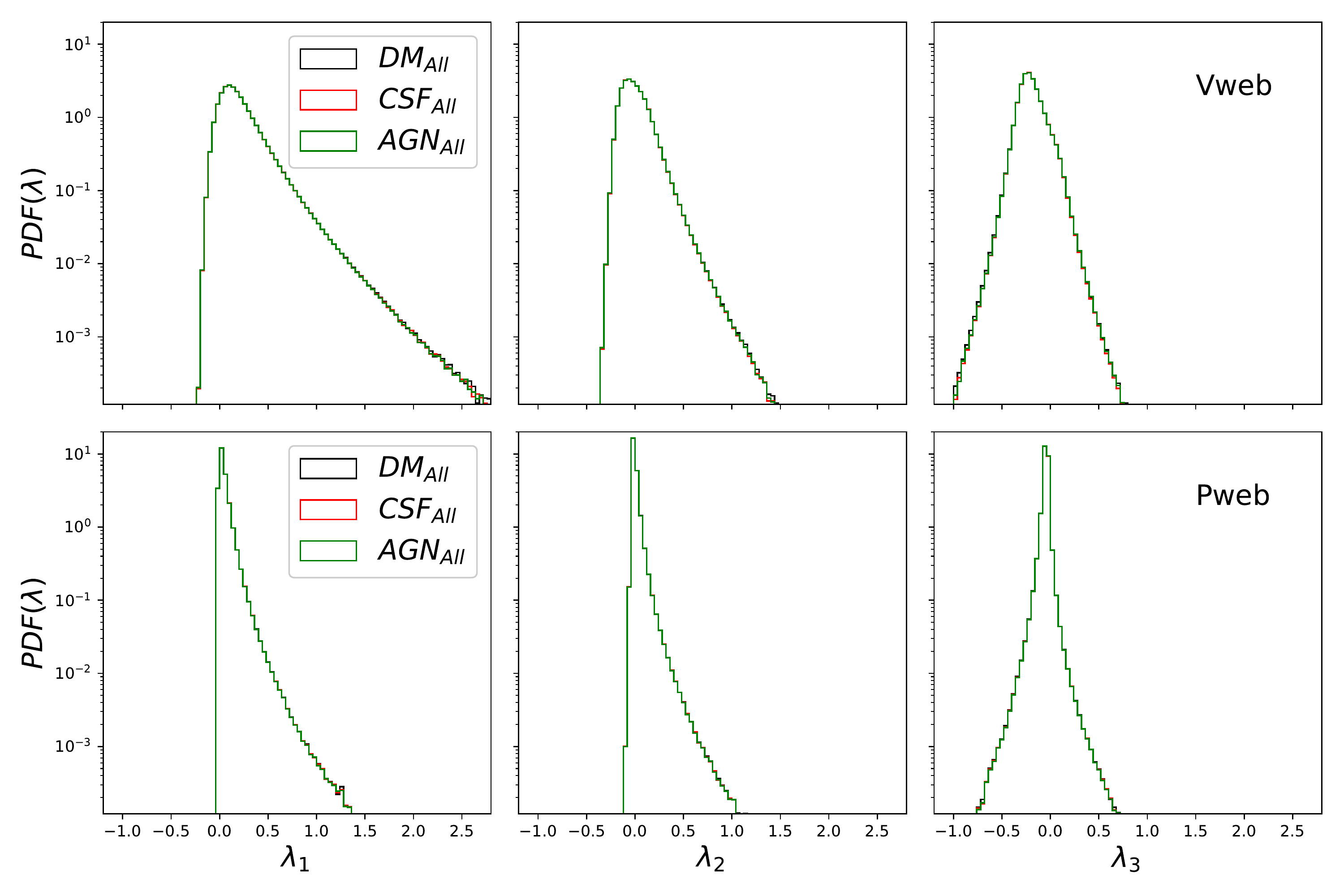}
 \caption{
  The probability distribution function of the three $\lambda$ parameters from the {\sc Vweb} code (upper panels) and from the {\sc Pweb} code (lower panels). As indicated in the legend, different colour histograms indicate different simulation runs.}\label{fig:lambda}
\end{figure*}
The two plots just presented and showing the cosmic web classification in a slice of the simulation have qualitatively shown that the modelling of sub-grid physics has little if any effect on it. Here we are now quantifying this by examining the probability distribution of the eigenvalues across each simulation. The result can be viewed in Fig.~\ref{fig:lambda}: From left- to right-hand panels, we show the distributions functions of $\lambda_{1,2,3}$. The top row shows the $\lambda^V$ and the lower row the $\lambda^P$ values and the different colour lines in each panel are for different simulation runs has indicated in the legend. It is very clear that the baryonic processes have little impact on the values of these three $\lambda$ parameters for both methods.

\begin{figure*}
 \includegraphics[width=\textwidth]{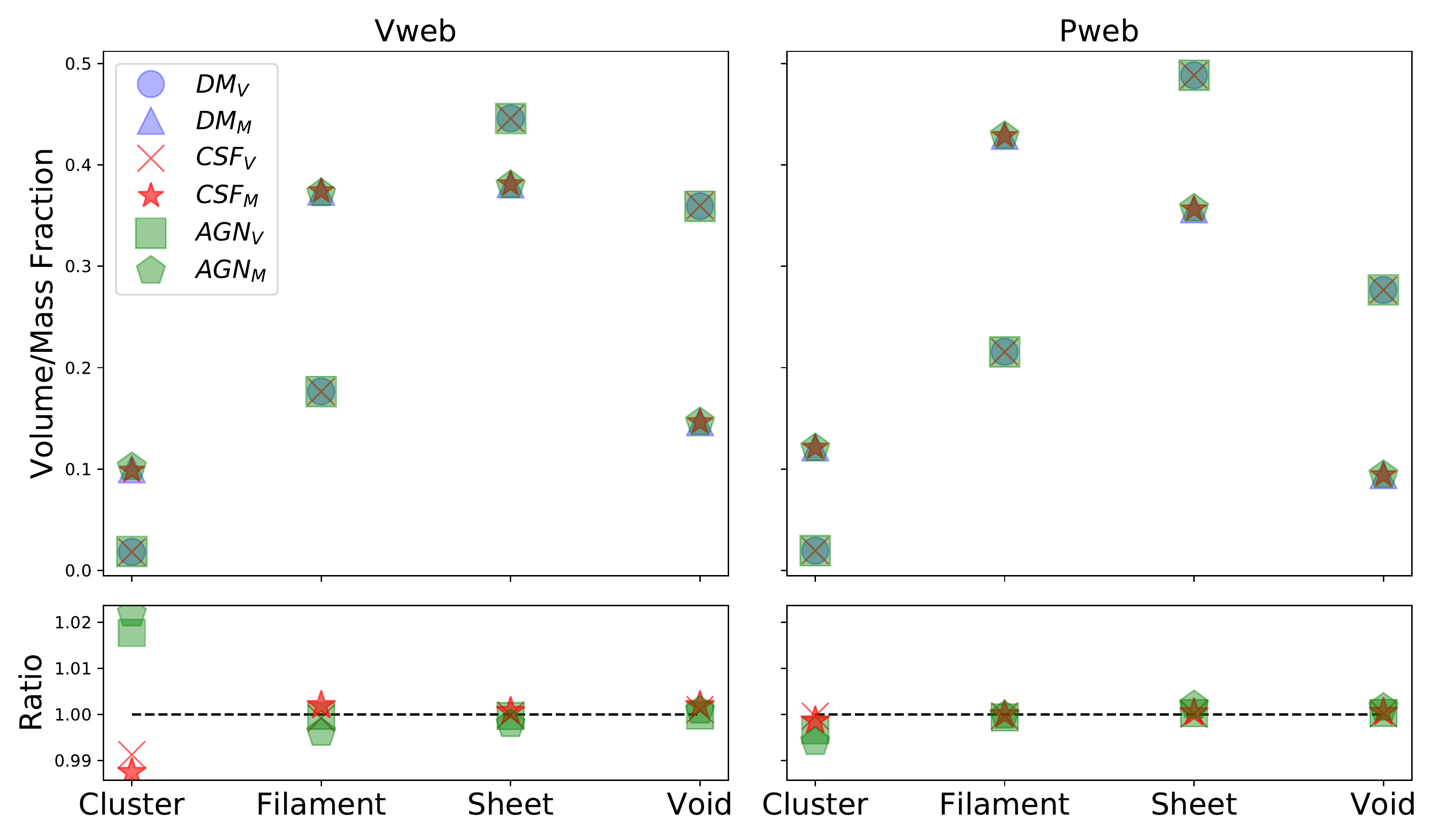}
 \caption{
  Upper panel: the volume and mass fractions of the structures. Lower panel: respective ratios between the CSF/AGN and DM-only runs. Left-hand column shows the results from the {\sc Vweb} code, while the right-hand column is for the {\sc Pweb} code. The meaning of the symbols is shown in the legend on the top left-hand panel, where sub-indexes V and M indicate volume and mass fractions.}\label{fig:be}
\end{figure*}

We further measure the influence of baryons by calculating the fraction of cells classified as knot, filament, sheet and void: In the upper panel of Fig.~\ref{fig:be}, we show both the volume fraction (indicated by subindex V in the symbol names) and the mass fraction (indicated by subindex M) for cells of a given classification (which is listed on the x-axis). {\sc Vweb} results are shown in the left-hand column, while right-hand column is for the results from the {\sc Pweb} code. As shown in the legend on the top left-hand panel, different colour symbols represent different runs. It is not surprising to see that their volume fractions are different to their mass fractions. That is because the mean density decreases from knot to void regions. For both methods, we see excellent agreement between these three simulation runs for both volume and mass fractions with variations lower than 2 per cent.

The variations are better seen in the lower panel of Fig.~\ref{fig:be}, which shows the quantitative difference between the respective fractions runs with respect to the DM-only run. For the {\sc Vweb} results, the AGN run tends to have both a slightly larger ($\sim$2 per cent) volume and mass fraction in the highest density region (i.e. knots), while the CSF run gives $\sim$1 per cent lower volume and mass fractions than the DM-only run. Without AGN feedback, the CSF run tends to have more concentrated knots with a relatively weaker velocity field, which tends to occupy less spatial volume and mass; while the strong AGN feedback not only stops star formation but also pushes matter into outer regions, which results in a large volume with a higher velocity field \citep[see discussion in][]{Ragone-Figueroa2013, Cui2014b, Cui2016a}. In filament, sheet and void regions, both mass and volume fractions show almost no change between these three runs. As the PWEB code directly uses the second derivatives of the potential, which are directly connected to the density via Poisson's equation, to classify these structures, there is even less difference ($\lesssim$1 per cent) between the two hydrodynamical runs and the DM simulation for all structures.

\subsection{The dark and gas components} \label{dark-gas}

\begin{figure*}
 \includegraphics[width=\textwidth]{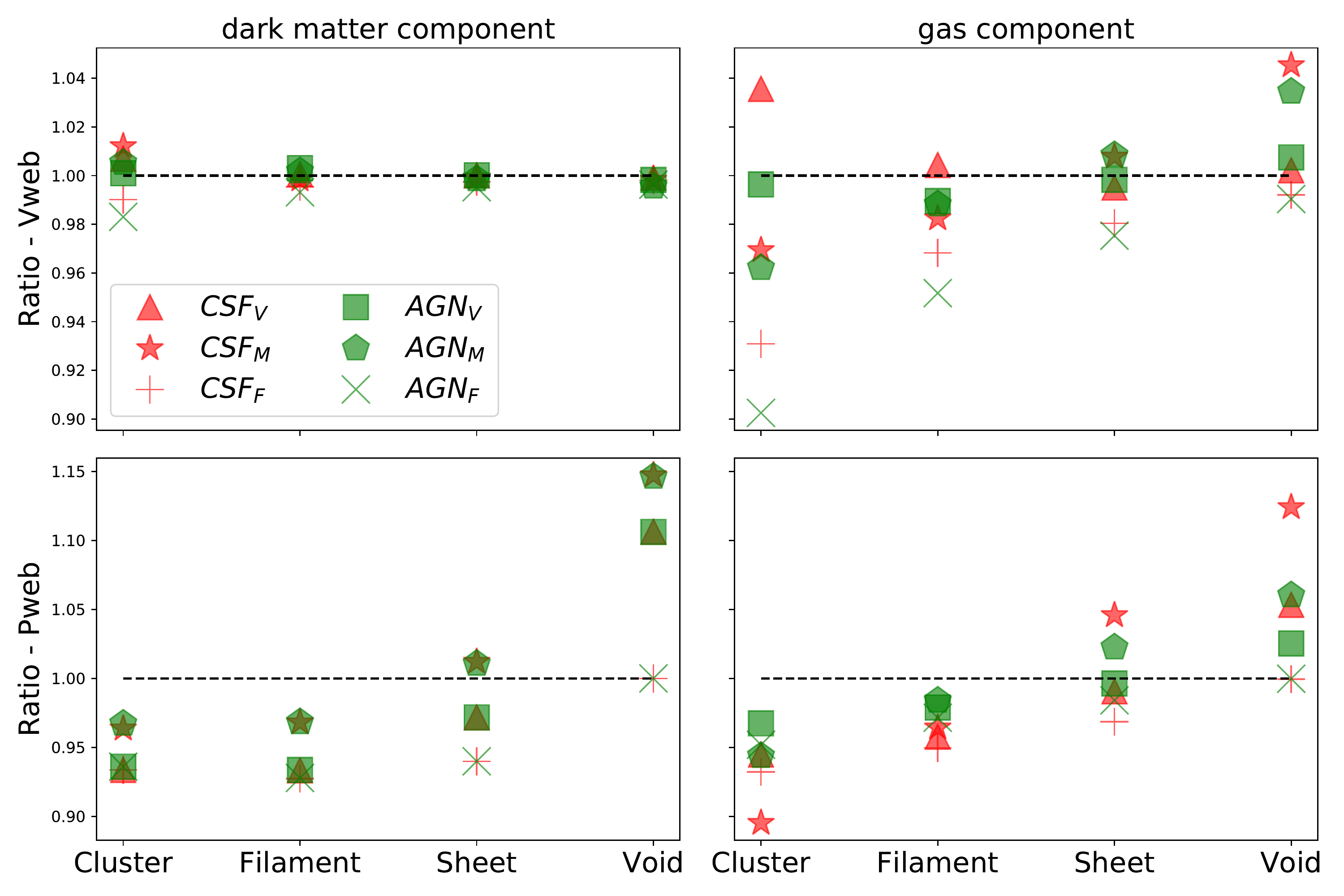}
 \caption{
  Left-hand panel: The relative fraction difference between the DM component and all matter; Right-hand panel: The relative fraction difference between the gas component and all matter. Upper row shows the {\sc Vweb} results, while bottom row is from the {\sc Pweb} code. Similar to Fig.~\ref{fig:be}, these sub-indexes V and M indicate volume and mass fractions, while the sub-index F indicates the consistency fraction for these mesh grids.}\label{fig:cf}
\end{figure*}
As shown in Figs.~\ref{fig:show_v} and \ref{fig:show_p}, both DM and gas components present a very similar cosmic web classification which in both cases agrees very well with the classification for all matter combined. Here, we now quantify this relation in more detail. To this extend we show in Fig.~\ref{fig:cf} the relative difference for both mass and volume fractions between the results using a single component to classify the cosmic web (DM on the left-hand panel and gas on the right-hand panel) and the results from all matter. The upper row shows the results from the {\sc Vweb} code, while the bottom row plots are the {\sc Pweb} results. As indicated in the legend on the top left-hand panel, we present these ratios with different colour symbols. Symbol names with subindex V indicate volume fraction, M means mass fraction and F is for consistency fraction. The consistency fraction is defined as $N_{i,c}/N_{i,a}$, where $i$ stands for knot, filament, sheet or void, $N_{i,a}$ is the grid number for the $i$th structure from all matter and $N_{i,c}$ is the consistent grid number, which is identified as the $i$th structure in both all matter and single components.

As shown in the top left-hand panel of Fig.~\ref{fig:cf}, the mass, volume and consistency differences between the results from DM and from all matter are very small ($\lesssim 1$ per cent) for the {\sc Vweb} code. This indicates that the velocity field generated by the DM is almost identical to the total velocity field. As DM dominates the velocity field at large scales, it is not surprising to see such agreement. The contributions from baryons (and their sub-grid modelling) to the velocity field are small, mainly at non-linear scale, which is smoothed out with little impact on the large-scale velocity field. This is in agreement with Fig.~\ref{fig:be} where we have seen that baryons have little effect on the cosmological structures. When using the gas only to classify the cosmic web, we find a slightly larger disagreement with the results from all matter:
\begin{itemize}
  \item[1.]  In knots, only $\sim$90 per cent of the cells in the AGN run are consistent with the results from all matter ($\sim$93 for the CSF run). And  the mass fraction for both the CSF and AGN runs are $\sim$4 per cent lower. The knot volume fraction in the CSF run is about $\sim$4 per cent higher, while the AGN volume fraction shows almost no change.
  \item[2.] For both filament and sheet regions, the changes are basically within 2 per cent for both CSF and AGN runs with over 96 per cent mesh grids cross-identified as the same environment.
  \item[3.] In void regions, the volume and consistency fractions are basically within 1 per cent for both runs. However, the mass fraction tends to be $\sim$4 per cent higher for both CSF and AGN runs.
\end{itemize}
These differences agree well with the expectations and the influence of baryonic processes: In knots, baryon processes are more violent. Thus, the velocity field derived from gas can deviate from the total velocity field, which results in a relatively lower mass fraction and about 10 per cent inconsistency fraction for both the CSF and AGN runs. In void environment, gas tends to have higher mass fractions although about 99 per cent of these void cells are in agreement with the all matter results. This indicates a slightly higher gas density contrast in the void regions compared to the DM one.

The relative differences for the {\sc Pweb} results are shown in the bottom panels. The ratios shown in bottom left-hand panel between the DM results and all matter results are much larger than the results from the {\sc Vweb} code. This means that the DM density field is more easily affected by baryon processes than the velocity field. This is not surprising because the velocity field is more sensitive to large-scale density fluctuation models. But the ratios shown for the gas component classification in the bottom right-hand panel are very similar to the results from the {\sc Vweb} code. This implies consistent changes of the gas density and velocity fields.

With both {\sc Vweb} and {\sc Pweb} codes, the filament and sheet structures identified with the gas component show very little disparity ($\lesssim 5$ per cent) with the results from all matter. This implies that gas filaments are almost identical to the DM filaments in velocity fields and are linearly proportional to the DM filaments in density fields. Since gas density is roughly proportional to the electron density, one can make a assumption on the magnetic field to get the synchrotron emission by combining its velocity field, which indicates the similar filamentary structure. In addition, although the $H_I$ gas condense mainly inside galaxies \citep[e.g.][through theoretical modelling]{Duffy2012,VN2016}, \cite{Takeuchi2014,Kooistra2017} found that it is sufficient to make the detection of the IGM gas in filaments through the $H_I$ 21-cm line emission. Therefore, the consistency between DM and gas filamentary structures theoretically allows us to precisely map the DM distribution in filaments with next-generation radio telescopes.

\section{Discussion and Conclusions}
\label{concl}

Using a series of cosmological simulations, which include different versions of baryonic processes, we study the distribution of matter at LSE. Three sets of simulations allow us to explore how baryons affect the LSE: (1) a dark-matter-only run as a gauge, (2) a simulation that includes gas cooling, star formation and supernova feedback, and (3) a run that additionally models AGN feedback. To all these simulations we applied both a velocity ({\sc Vweb}) and potential ({\sc Pweb}) based cosmic web classification on a regular grid of physical spacing $\sim 1.6 \Mpc$. Both methods assign to each cell a tag `knot', `filament', `sheet' or `void' that serves as the web classifier.

\medskip

\noindent
In this paper we focus on two simple questions:

\begin{itemize}
 \item How do baryon processes affect the cosmic web classification?
 \item How well can gas be used as a tracer of the cosmic web?
\end{itemize}

\noindent
Main results are summarized as follows
\begin{enumerate}
 \item As expected, baryons have a very weak impact on the mass and volume fractions of cosmological structures. The difference is $\lesssim 2$ per cent for the knot and void regions. There is almost no difference for filament and sheet regions.
 \item We confirm that the gas component of these simulations follows the same cosmic web as the total and DM component alone, respectively -- at least for filament and sheet regions; it provides an un-biased tracer of the cosmic web to be used with large-scale surveys such as the SKA.
 \item Although the volume and mass fractions of each cosmological structure show several per cents of differences between the {\sc Vweb} and {\sc Pweb} codes, our main results, which present in ratio, are very similar between the two codes and less susceptible to the sample variance and numerical artefacts.
\end{enumerate}

As expected and also agreed with previous findings, the baryons leave a very weak impact on these cosmological structures. Because the DM dominates in large-scale, while baryonic processes mainly happen in high non-linear scale -- galaxy and galaxy clusters. It is very hard for these non-linear changes to influence the linear scale structures. By comparing the dark-matter-only simulation with its hydro-dynamical runs, \cite{VanDaalen2011} \citep[see also][for similar results]{Rudd2008} found less than $\sim 1$ per cent of power spectrum change at $k \leq 0.3 h Mpc^{-1}$. In agreement with our conclusion, \cite{Zhu2017} also found that the mass differences between baryonic and DM-only runs are very small at $z = 0$ for filaments. However, they reported a much larger difference for the other three cosmological structures types (see details in their fig. 7). We have checked our result with a larger $\lambda^V_{th} = 0.3$ for the {\sc Vweb} code, which shows a consistent result for baryonic effect with $\lambda^V_{th} = 0.1$. Therefore, the disagreement between theirs and our results may be caused by their small simulation volume.

As shown by \cite{Dolag2006}, filamentary structures are very hard to detect at several wavelengths, for example, X-ray and thermal Sunyaev-Zel'dovich. But detecting these filaments structures through radio emissions with the next-generation radio telescopes (e.g. SKA) is very promising \citep[e.g.][]{Vazza2015, Kale2016, Brown2017}. Using both the {\sc Vweb} code and the {\sc Pweb} code, we have shown that the gas component can be used as a tracer for these cosmological structures, especially for filaments. These two methods use gas velocity ({\sc Vweb}) and density ({\sc Pweb}) to classify LSE. Both quantities are theoretically connected with radio signals, such as the synchrotron or the $H_I$ 21-cm line emission, which raise new possibilities for detecting gas filaments. Interestingly, using a shock-finding algorithm \citep{Planelles2013}, \cite{Martin-Alvarez2017} studied shocks in different environments and found that shocks in filaments display a spectral slope coincident with the gas matter power spectrum. That implies a consistency distribution of gas in both real and phase spaces. Using a method for estimating the radio emission, \cite{Araya-Melo2012b} predict that the radio flux of filaments at redshift $z \sim 0.15$, and at a frequency of 150 MHz, should be $S_{150 MHz} \sim 0.12 μJy$. However, the detailed connections between gas properties and detected radio signals relies on the theoretical model for estimating the radio emission. Further investigation is required.

Since these properties of LSEs are insensitive to the uncertain knowledge of baryonic processes, they can, in principle, be used to constrain cosmological models. However, this remains to be seen how we can extract these {\sc Pweb} and {\sc Vweb} diagnostics from observational data. Besides the commonly used methods such as power spectrum \citep[e.g.][]{Cui2010}, halo mass function \citep[e.g.][]{Cui2012b} and void regions have been suggested for constraining cosmological models \citep[e.g.][]{Cai2015,Sutter2015}. However, baryons may leave an effect on the void structures and bias the result. Our result confirms that these void regions, especially their volume fraction, are less affected by baryonic processes. Therefore, the finding in \cite{Cai2015} that the void abundances in $f(R)$ gravity can differ significantly from in GR, can be directly used to constrain cosmology models with the observed voids, which can be detected with gas as the tracer.

Although the fractional properties of these cosmological structures are very similar for the simulations used here, the detailed mass distributions inside each structure may have changed due to the baryonic processes. In a follow paper we plan to take a closer look at the objects that are found in the various cosmic web types and compare their differences amongst the simulations and how they have been influenced by the baryon physics.

\section*{Acknowledgements}

We thank the referee, Dr. Elmo Tempel, for his useful and constructive report.
The authors would like to thank Youcai Zhang for useful discussions and thank Giuseppe Murante for preparing these simulations.
This work has made extensive use of the {\sc Python} packages --- {\sc Ipython} with its Jupyter notebook \citep{ipython}, {\sc NumPy} \citep{NumPy} and {\sc SciPy} \citep{Scipya,Scipyb}. All the figures in this paper are plotted using the python {\sc matplotlib} package \citep{Matplotlib}. This research has made use of NASA's Astrophysics Data System and the arXiv preprint server. Simulations have been carried out at the CINECA supercomputing Centre in Bologna, with CPU time assigned through ISCRA proposals and through an agreement with the University of Trieste. SB and GM acknowledge support from the INDARK INFN grant and 'Consorzio per la Fisica di Trieste'.

XY is supported by the national science foundation of China (grant Nos. 11233005, 11621303)

WC, AK and GP are supported by the {\it Ministerio de Econom\'ia y Competitividad} and the {\it Fondo Europeo de Desarrollo Regional} (MINECO/FEDER, UE) in Spain through grant AYA2015-63810-P as well as the Consolider-Ingenio 2010 Programme of the {\it Spanish Ministerio de Ciencia e Innovaci\'on} (MICINN) under grant MultiDark CSD2009-00064. AK also acknowledges support from the {\it Australian Research Council} (ARC) grant DP140100198. He further thanks The Jam for a town called malice.

\bigskip

\bibliographystyle{mnras}
\bibliography{cosmic-web}
\bsp

\appendix

\section{resolution effects}
\label{A:denspeak}
\begin{figure*}
 \includegraphics[width=\textwidth]{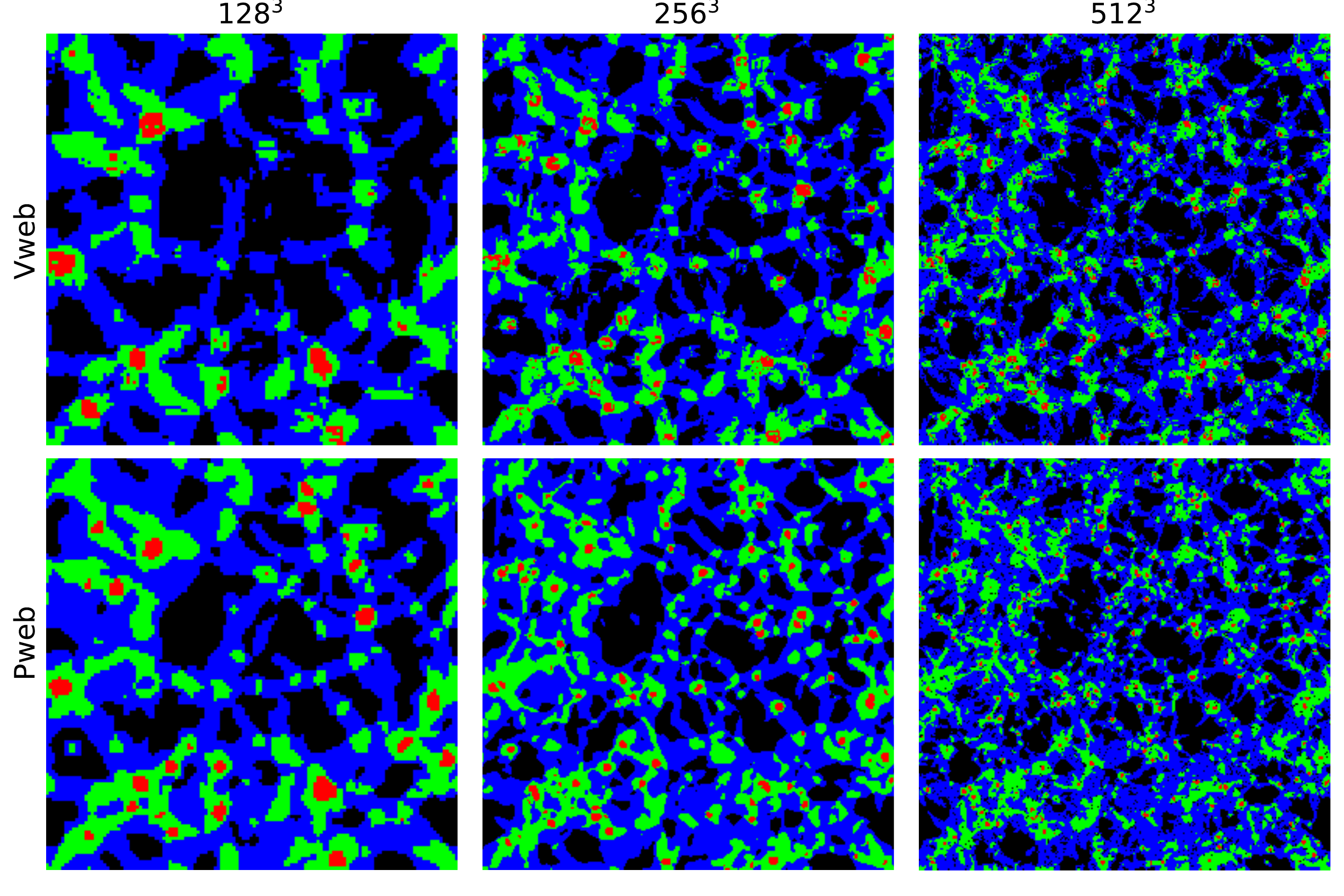}
 \caption{
  Similar to Fig.~\ref{fig:show_v}, an illustration of the projected cosmology structures. Knots, filament, sheet and void regions are shown in red, blue, green and black, respectively. From left- to right-hand side, the three columns show the results from different mesh grid numbers shown in the titles. Upper panels show the results from the {\sc Vweb} code, while lower panels show the results from the {\sc Pweb} code.}\label{fig:rs_show}
\end{figure*}

\begin{figure*}
 \includegraphics[width=\textwidth]{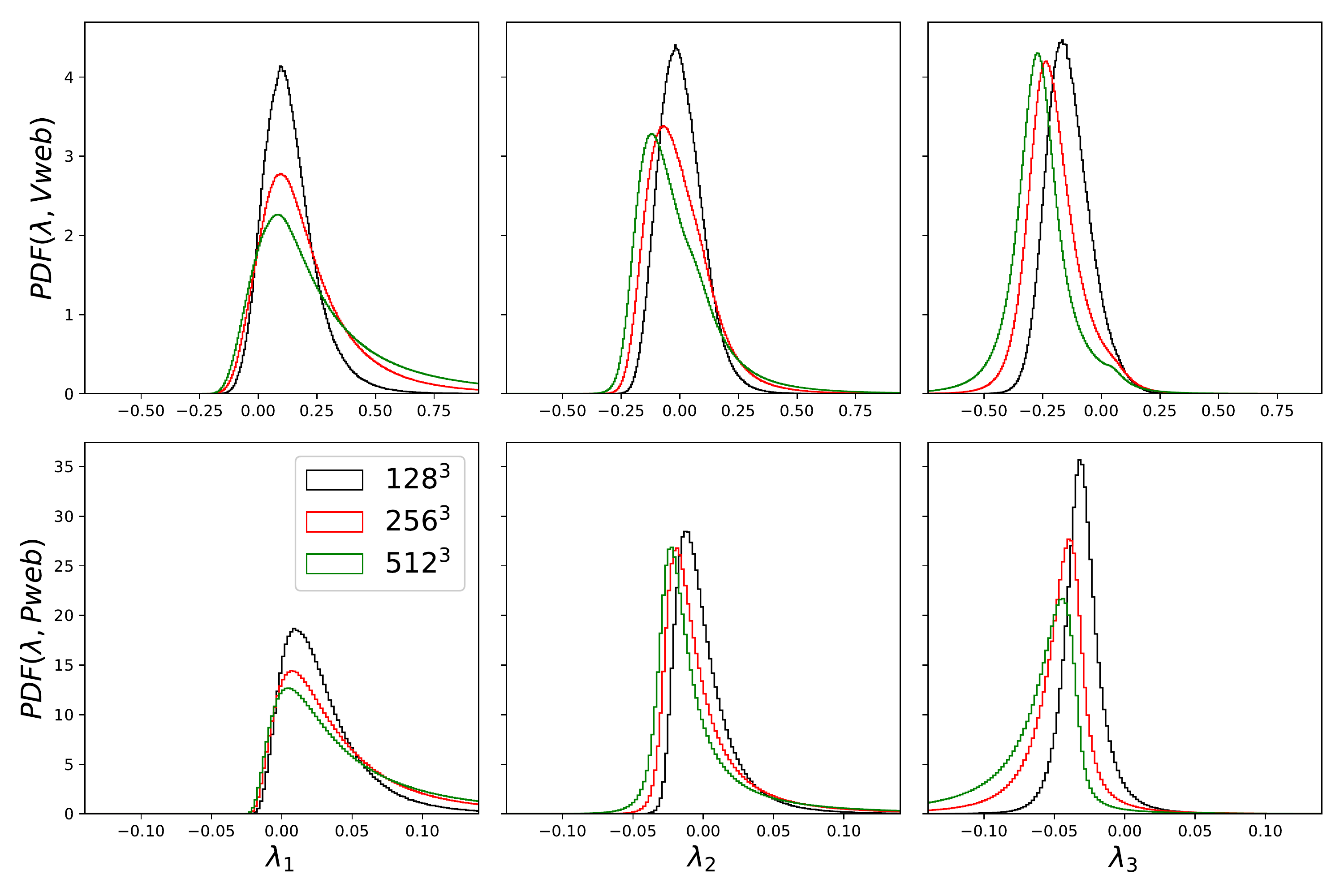}
 \caption{
  Similar to Fig.~\ref{fig:lambda}, probability distribution of the lambda parameters. Different smoothing lengths are shown with different colours. The three $\lambda$ parameters are shown from left- to right-hand columns. While the results from {\sc Vweb} and {\sc Pweb} are shown in up and bottom rows, respectively.}\label{fig:rs_pdf}
\end{figure*}

\begin{figure*}
 \includegraphics[width=\textwidth]{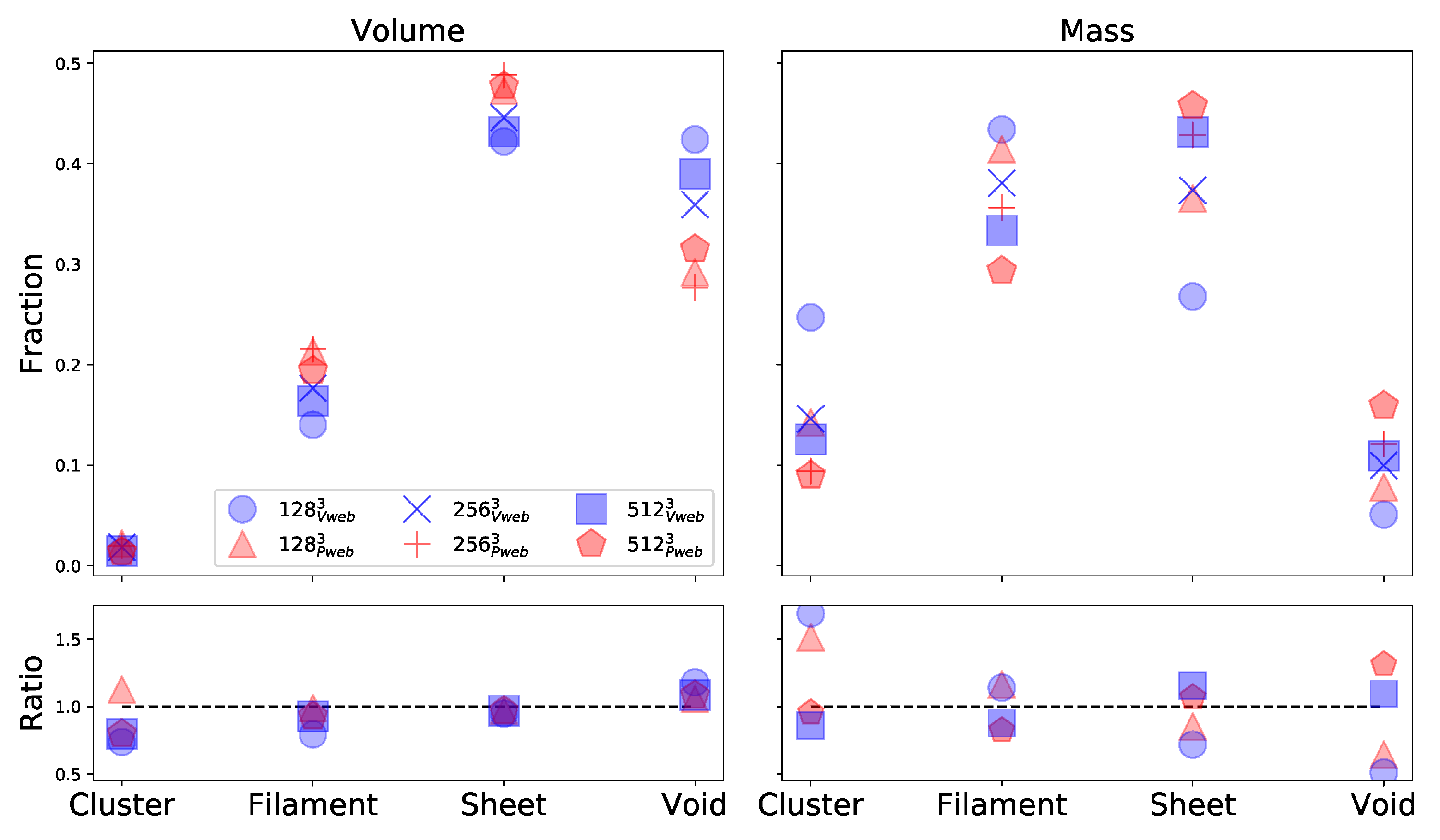}
 \caption{
  The volume (left-hand panel) and mass(right-hand panel) fractions of the different cosmology structures. Relative differences to the $256^3$ mesh number are shown in the second row. Refer to the legend in the top left-hand panel for the meanings of different symbols.}\label{fig:rs_fraction}
\end{figure*}

Both {\sc Vweb} and {\sc Pweb} have several parameters that can impact their results. The most important one is the $\lambda_{th}$, which has been tested in \cite{Carlesi2014}. They showed that $\lambda_{th} = 0.1$ gives a good visual match to the density field (see also the Fig.~\ref{fig:show_v} in this paper). Therefore, we fix $\lambda_{th} = 0.1$ for the {\sc Vweb} code. The $\lambda_{th}$ for the {\sc Pweb} code is chosen by mimicking the cosmological structures from the {\sc Vweb} code. Even this $\lambda_{th}$ has an impact on our results (which should be unlikely because our main results are shown in ratio), we do not explore these improper values of the $\lambda_{th}$, which will give unphysical structures. Instead, we focus on the other parameter --- mesh grid number at this appendix. Here, we only use the DM-only run to examine its effects.

Similar to Fig.~\ref{fig:show_v}, Fig.~\ref{fig:rs_show} illustrates the distribution of the cosmology structures (indicated by different colours) from different codes (up row: {\sc Vweb}; bottom row: {\sc Pweb}) and different grid numbers (left-hand column: $128^3$; middle column: $256^3$; right-hand column: $512^3$). Each plot shows the 2D project of the simulation box size with a thickness of $\sim 3.2 \Mpc$ for $128^3$, $\sim 1.6 \Mpc$ for $256^3$ and $\sim 0.8 \Mpc$ for $512^3$ mesh numbers. We do not use higher mesh numbers at here for the examinations, because too small mesh cell can not provide enough statistics for these structures. Visually, finer mesh shows the structures in more detail, but the structures looks also more noisy. As expected, the cosmology structures become smoother as the mesh numbers decreases. Therefor, we adopt $256^3$ mesh number for the analysis in this paper. We will loose detailed information of the cosmological structures with large mesh grid size. While too small grid size will result in noise and improper structure shapes, which could be caused by the less sample particles in our simulation. For all three mesh numbers, {\sc Vweb} and {\sc Pweb} show consistent results for different cosmological structures.

Similar to Fig.~\ref{fig:lambda}, we show probability distributions of these lambda parameters. Up row shows the results from the {\sc Vweb} code, while bottom row shows the results from the {\sc Pweb} code. From left to right-hand columns, we show the three $\lambda$ parameters with different mesh grid numbers, which are presented by different colours. The overall impression for these plots is that the mesh grid number has little effect on the three $\lambda$ parameters by means of their peak positions and distribution shapes. The lowest mesh grid number $128^3$ tends to have a higher value of the peak for all three $\lambda$ parameters in both methods. While the probability distributions from $256^3$ grid number is much closer to the $512^3$ grid number, this shows a sign of convergence.

The volume (left-hand column) and mass (right-hand column) fractions for the different mesh grid numbers and from two methods are shown in Fig.~\ref{fig:rs_fraction} with different colourful symbols, of which their meanings are indicated in the legend of the top left-hand panel. Similar to Fig.~\ref{fig:be}, $x$-axes show the different cosmology structures. We show the relative differences with respect to $256^3$ grid number in second row. The volume fraction ratios show relatively small change for all four cosmology structures and for both methods. However, the mass ratios between the two mesh grid numbers ($128^3$ and $256^3$) show a clear decreasing effect from $\sim 1.6$ at knots to $\sim 0.6$ at voids, while the difference between $512^3$ and $256^3$ is relatively small ($\lesssim 10$ per cent) for all four cosmological structures. This means the mass fraction is more sensitive than the volume fraction to the resolution effect with smaller grid size. Again, there is very small change between the two methods.

As our main results in this paper are shown in ratio, we expect that these resolution and parameter effects will be canceled and leave no effect on our results, especially for the filament structure fractions, on which these effects are basically within $\sim 10$ per cent as shown in Fig.~\ref{fig:rs_fraction}.

\label{lastpage}
\end{document}